\documentclass[aps,superscriptadddress,noshowkeys,showpacs,amsmath,amssymb,pra,reprint,letterpaper]{revtex4-1}
\usepackage{graphicx}
\usepackage{enumitem}
\usepackage{braket}
\usepackage[utf8]{inputenc} 
\usepackage[english]{babel}
\usepackage[title,titletoc,toc]{appendix}
\usepackage[usenames,table]{xcolor}
\usepackage[subnum]{cases}

\begin{document}
\title{A Quantum Extension of the Semiclassical Theory of Electrical Susceptibility}
\author{Jairo David Garc\'ia}
\affiliation{Instituto de F\'isica, Facultad de Ciencias Exactas y Naturales, Universidad de Antioquia UdeA; Cl 70 No. 52-21, Medell\'in, Colombia}
\affiliation{Dinntec SAS, jgarcia@dinntec.com.co}
\author{Boris A. Rodr\'iguez }
\affiliation{Instituto de F\'isica, Facultad de Ciencias Exactas y Naturales, Universidad de Antioquia UdeA; Cl 70 No. 52-21, Medell\'in, Colombia}

\begin{abstract}
It is shown here how the semiclassical theory of electrical susceptibility can be extended to the case in which both radiation and matter are quantized. This is done specifically for the cases of linear and second order susceptibilities. The expressions derived allow to determine a set of conditions for the validity of the semiclassical approximation and predict interesting new phenomena such as photon number dependent susceptibility. 


\end{abstract}

\maketitle

\section{INTRODUCTION}
The study of light-matter interactions can be addressed  by assuming that matter obeys the quantum laws of physics, while electromagnetic field obeys the classical ones. It can also be effected within a more sophisticated scheme, in which both matter and radiation fulfill quantum rules. We refer here to the former approach as the semiclassical treatment and to the later as the quantum treatment. It is widely known that semiclassical and quantum treatments have profound differences: the first is unable to describe important aspects of the observed dynamic of light-matter interactions, as vacuum Rabi oscillations; while the last is successful in describing such behaviors and predicting a number of new phenomena \cite{Aspect}. The quantization of the electric field is the best tool available to accurately account for the role of the quantum states of light, as can be seen, for example, in the collapses and revivals in the Jaynes–Cummings model, with the field initially in a coherent state \cite{Eberly} and in the creation of entangled photon pairs in parametric down conversion \cite{Burnham, Shih}.
  
Analytical expressions for electrical susceptibilities of different orders are well known in the semiclassical regime, where susceptibility is assumed to be an attribute of each material. Those expressions have been used for a number of applications, from understanding the susceptibility dependence on medium properties and predicting its possible values \cite{Boyd, Shen}, to designing quantum dots (QD) based optical amplifiers \cite{Sugawara} and heterostructures with enhanced nonlinear responses \cite{Gurnick, Rosencher}. Nonetheless, with increasing interest in light-matter interaction in micro-cavities, where quantum features of light become more relevant, we are prone to exceed the validity scope of those expressions. 

The purpose of this paper is to extend the semiclassical theory of electrical susceptibility using the second quantization of electromagnetic field. This extension should not only help us to decide when it is safe to use the semiclassical approach and when quantum states of radiation modify susceptibility, but, in the future, it should provide more information about processes such as the use of QD to produce nonclassical states of light \cite{Faraon, Vahala, Gerry} and the absorption and emission of light by QD embedded in micro-cavities. Few efforts have been made in developing such extension: Voon and Ram \cite{Voon} derived an expression for the second order susceptibility in III-V and II-VI semiconductor heterostructures, using a fully quantum mechanical theory of the electron-photon interaction; Andrews et al. \cite{Andrews} did it for crystals using the S-Matrix method. Both works use at some point in their procedures the semiclassical idea of the susceptibility as an inherent property of the medium. Even very comprehensive works on nonlinear quantum optics such as Drummond’s \cite{Drummond}, use expressions that have no way to account for the impact of the quantum state of light in the values of susceptibilities.  

Here we compare the expectation values of polarization, expressed, on one side, as a power series of electric field, and on the other, as a dipole moment. Density matrix elements in both expectation values are found via a perturbative solution of the master equation, and the values of susceptibilities are deduced by equating terms of the same order in both series for polarization. The extension is achieved by using a quantized electric field instead of a classical one and allowing the density matrix to include information about the quantum state of light. By using this approach, we derive expressions for the first $ \left(\chi ^{\left(\omega\right)}\right) $ and second order susceptibilities, specifically for the case of second harmonic generation -SHG- $ \left(\chi^{\left(2\omega\right)}\right)$, although the procedure can be extended to any case or order. According to the expressions found, the value of susceptibility depends on the state of light. This has a clear implication: susceptibility is not exclusively a property of the medium but of its interaction with light. Ignoring this fact amounts to introduce a semiclassical approximation. An interesting result of this work is a set of conditions for the validity of such approximation.   

This paper is organized as follows: in Sec. \ref{HS} we delimit the system under analysis, and the relevant approximations done, paying special attention to quantum nature of light. In Sec. \ref{TES} we present the dynamical equations to be solved and sketch the perturbative solution method. The results for the zeroth, first and second order terms of the expansion are presented in Secs. \ref{SX0}, \ref{SX1}  and \ref{SX2}, respectively. In Sec. \ref{SAQD} we present an application of the theory for an asymmetric QD. In Sec. \ref{SDC} we discuss the results and summarize conclusions. 

\section{THEORY}
\label{THEORY}
\subsection{Model}
\label{HS}
Let us consider an open quantum system comprised by a single radiation mode and an active medium. Using the dipolar approximation \cite{Aspect}, the Hamiltonian $\hat{H}$ of the system can be written as the sum of the free contributions of the matter and field, and the dipolar interaction Hamiltonian:
\begin{equation}
\label{H}
\hat{H} = \hat{H}_{\mathrm{mat}} + \hat{H}_\omega + \hat{H}_{\mathrm{int}}, 
\end{equation} 			 
with:
\begin{equation}
\label{mat}
\hat{H}_{\mathrm{mat}} = \sum\limits_{s}{\varepsilon_{s} \left|s \right\rangle  \left\langle s \right|} ,      
\end{equation}    
\begin{equation}
\label{Hrad}
\hat{H}_\omega =\hbar \omega \hat{a}^\dag \hat{a}, 
\end{equation}
\begin{equation}
\label{Hint}
\hat{H}_{\mathrm{int}}=-\hat{\boldsymbol{\mu}}\cdot \bf\hat{E}. 
\end{equation}
$\varepsilon_{s}$ and $\left|s \right\rangle $ are respectively the eigenvalues and eigenvectors of the free mater Hamiltonian $\hat{H}_{\mathrm{mat}}$, $\hat{a}^\dag $ and $\hat{a}$ are the creation and annihilation operators of the field mode with frequency $\omega$, ${\bf \hat{E}}$ is the electric field operator and $\hat{\boldsymbol{\mu}}$ is the electric dipole moment operator of the medium. In semiclassical theory, the dipole moment is assumed to have null diagonal terms, we shall study the general case in which the active medium may be subject to asymmetric external fields \cite{Kibis} or have a peculiar shape that breaks spatial symmetry \cite{Jacak}; thus, we will allow diagonal terms of the dipole matrix to take nonzero values. 

For the sake of simplicity, we use standard second quantization in free space, thus, the electric field operator of the mode confined to a cavity of volume $V$, polarized along direction $\mathbf{e}$, using the Coulomb gauge, is given by \cite{Aspect}:
\begin{eqnarray} 
\label{E}
{\bf \hat E} &=& i \left( \cfrac{\hbar \omega}{2 \epsilon_0 V} \right)^\frac{1}{2} {\bf e} \left( \hat{a} e^{-i\omega t} - \hat{a}^\dag e^{i\omega t}\right) 
\\
&=& {\bf e} \left( \hat E^{\left( \omega  \right)}e^{- i\omega t} + \hat{E}^{\left( -\omega  \right)}e^{i\omega t} \right)= \hat{ \tilde{\bf{E}}}^+ + \hat{ \tilde{{\bf E}}}^-, 
\nonumber
\end{eqnarray} 
where $ \epsilon_0$ stands for the vacuum permitivity.

We may describe the quantum mechanical state of the system by the means of the density matrix:
\begin{eqnarray} 
\label{rho}
\hat{\rho} &=& \sum\limits_{kn,lm}{{{\rho }_{kn,lm}}} \ket{k} \otimes \ket{n}  \bra{l} \otimes \bra{m}
\\
&=& \sum\limits_{kn,lm}{{{\rho }_{kn,lm}}} \ket{k,n} \bra{l,m} . 
\nonumber
\end{eqnarray} 
Labels $k$, $l$ stand for matter states; labels $n$, $m$ stand for Fock states of the field. 

\subsection{Perturbative approach}
\label{TES}
The time evolution of the matter-field state is governed by the master equation:
\begin{equation}
\label{maestra}
\frac{{\partial \rho_{kn,lm}}}{{\partial t}} =  - \frac{i}{\hbar}\left[ {\hat{H},\hat{\rho} } \right]_{kn,lm}  - \gamma_{kn,lm} \left( \hat{\rho}- \hat{\rho}^{\mathrm{ss}} \right)_{kn,lm},
\end{equation}                           
where, as customary \cite{Boyd, Rosencher}, we have added a phenomenological damping term $\gamma_{kn,lm} \left( \hat{\rho} - \hat{\rho}^{\mathrm{ss}} \right) $ to take into account non-Hamiltonian processes such as dissipation, decoherence and/or pumpings that push the system, at a rate $\gamma_{kn,lm} $, towards a steady state $\hat{\rho}^{\mathrm{ss}}$. To be more rigorous, we should use another kind of master equation (for example, in the Born-Markov regime, a Lindblad master equation should be used); we prefer the phenomenological equation (\ref{maestra}), as accustomed in the semiclassical theory of susceptibility, to allow a more straightforward comparison between the two approaches. This election implies the appearance of the rates $\gamma _{kn,lm}$ that carry a lot of information about the system-environment interaction: the details of all the complex processes and mechanisms that should be described in the more general framework of open quantum systems \cite{Petruccione}. This imposes a limitation on our treatment because such information is not readily available by other means and the rates have a major impact on the values of susceptibilities. As a matter of fact, some of the main differences between the semiclassical theory and this quantum extension, arise from the possible values of such rates, that, in this case, depend both on matter and field states.

In order to solve (\ref{maestra}), we assume the interaction to be modulated by a perturbation parameter $\lambda $, and seek the correspondent perturbed states of the system. This amounts to replace ${\bf \hat E}$ by $\lambda {\bf \hat E}$ and to expand the state of the system as a power series of $\lambda $, wherever they appear: 
\begin{equation}
\label{rhopert}
\rho _{kn,lm} = \sum\limits_{s = 0}^\infty \lambda ^s\rho _{kn,lm}^{\left( s \right)}.
\end{equation}  			         
We will require $\rho_{kn,lm}$, but not each of the terms $\rho_{kn,lm}^{( s)}$, except $\rho_{kn,lm}^{(0)}$, to satisfy the conditions imposed on density matrices. Besides, 
(\ref{rhopert}) shall be a solution of equation (\ref{maestra}) for any value of $\lambda $. This condition holds if the coefficients of each power of  $\lambda $ satisfy (\ref{maestra}) separately, hence, the following set of relations must be valid:
\begin{eqnarray}
\label{drho0}
\cfrac{{\partial \rho _{kn,lm}^{\left( 0 \right)}}}{\partial t} =  &-& i\omega _{kn,lm} \rho _{kn,lm}^{(0)} 
\\
&-& \gamma _{kn,lm}{\left( {\rho _{kn,lm}^{(0)} - \rho _{kn,lm}^{ss}} \right)}, 
\nonumber
\end{eqnarray}
\begin{eqnarray}
\label{drhos}
\cfrac{\partial \rho _{kn,lm}^{\left( s \right)}}{\partial t} =  &-& \left( i \omega _{kn,lm} + \gamma _{kn,lm} \right)\rho _{kn,lm}^{\left( s \right)}  
\\
&-& \cfrac{i}{\hbar }\left[ { \hat{H}_{\mathrm{int}}} ,\hat{\rho}^{\left( {s - 1} \right)} \right]_{kn,lm}, 
\nonumber
\end{eqnarray}
where $\omega _{kn,lm}$ is the transition frequency of the system: 
\begin{equation} 
\label{frectrans}
\omega _{kn,lm} = \omega _{kl} + \omega _{nm} = \hbar^{(-1)} \left( \varepsilon _k - \varepsilon_l \right) + \omega \left(n - m \right).
\end{equation}
Recurrence relations (\ref{drho0}) and (\ref{drhos}) determine contributions of different orders to density matrix. The zeroth order contribution is given by the equilibrium solution of (\ref{drho0}): 
\begin{equation}
\label{rho0rhoeq}
\rho_{kn,lm}^{\left( 0 \right)} = \frac{\gamma_{kn,lm}}{\gamma_{kn,lm} + i\omega_{kn,lm}} \rho_{kn,lm}^{\mathrm{ss}} = \rho_{kn,lm}^{\mathrm{eq}}.    
\end{equation}  			
This solution plays an important role as it will appear in higher order terms. In semiclassical theory, this equilibrium density matrix is assumed to be diagonal \cite{Boyd}, but we shall not abide by this because it is well known that non-diagonal states are common for the electric field and the matter. 

Once $\rho^{(0)}$ is known, we can integrate Eq. (\ref{drhos}) with $s = 1$ to find the first order contribution, $\rho^{\left( 1 \right)}$, which contains terms oscillating in time with frequencies $ \pm \omega$.
\begin{eqnarray}
\label{rho1}
\rho_{kn,lm}^{(1)} &=&  \rho_{kn,lm}^{( \omega)} e^{-i\omega t} + \rho_{kn,lm}^{( -\omega)} e^{i\omega t}
\\
&=& \tilde \rho_{kn,lm}^{(\omega)} + \tilde \rho_{kn,lm}^{(-\omega)}. 
\nonumber
\end{eqnarray}
The integration procedure is described in detail in standard textbooks \cite{Boyd} and, for the term $\tilde \rho _{kn,lm}^{\left( \omega  \right)}$, leads to:
\begin{eqnarray}
\label{rhow}
\tilde \rho _{kn,lm}^{\left( \omega  \right)} = &\cfrac{i}{\hbar}& \sum\limits_{LM} \left\{ \left\langle kn \right|\hat{\boldsymbol{\mu}} \cdot {\bf \hat{\tilde E}^+}\left| LM \right\rangle \rho_{LM,lm}^{\mathrm{eq}}  \right. 
\\
&-& \left.\rho_{kn,LM}^{\mathrm{eq}}\left\langle LM \right| \boldsymbol{\mu}  \cdot {\bf \hat{\tilde E}}^+ \left| lm \right\rangle \right\}\xi \left( \omega  \right), 
\nonumber          
\end{eqnarray}
where $L(M)$ runs over matter (field) states and:
\begin{equation}
\label{xi}
\xi \left( \omega \right)= e^{-i\omega t} \left[   \omega _{kn,lm} - \omega - i \gamma _{kn,lm} \right]^{-1}.
\end{equation}             
By inserting (\ref{E}) in (\ref{rhow}) we obtain:
\begin{eqnarray}
\label{rhowf}
\tilde \rho_{kn,lm}^{\left( \omega \right)} &=&  i
\left( \cfrac{\omega}{2\hbar \epsilon_0 V} \right)^{\frac{1}{2}} \sum\limits_{L \jmath} \left\{ \sqrt {n+1} \rho_{L(n+1),lm}^{\mathrm{eq}}\; \mu_{kL}^{\breve{\jmath}} \qquad \quad \right.  
\\ 
&& - \left. \sqrt {m} \rho_{kn,L(m-1)}^{\mathrm{eq}} \mu_{Ll}^{\breve{\jmath}} \right\} e_{\breve{\jmath}}\; \xi \left(\omega \right). 
\nonumber
\end{eqnarray}
The index $\breve{\jmath}$ runs over Cartesian coordinates and $e_{\breve{\jmath}}$ are the components of the polarization vector along direction $\breve{\jmath}$. An analog expression is found for $\tilde \rho _{kn,lm}^{\left(-\omega \right)}$. 
The second order contribution to the density matrix of the system, $\rho^{\left(2\right)}$, can be found by iteration of the process; it contains a constant term in time and terms oscillating in time with frequencies $\pm 2\omega$:

\begin{eqnarray}
\label{rho2}
\rho_{kn,lm}^{\left(2\right)} & = & \rho_{kn,lm}^{\left( 2\omega \right)}e^{-2i\omega t} + \rho_{kn,lm}^{\left(-2\omega \right)}e^{2i\omega t}+\rho_{kn,lm}^{\left(0\omega \right)} \qquad 
\\
&= & \tilde \rho_{kn,lm}^{\left( 2\omega \right)} + \tilde \rho_{kn,lm}^{\left(-2\omega \right)} + \rho_{kn,lm}^{\left(0\omega \right)}. \nonumber
\end{eqnarray}

\begin{widetext}
The term oscillating in time with frequency $ 2 \omega $ has the form:

\begin{eqnarray}
\tilde \rho_{kn,lm}^{\left( 2\omega\right)} &=&  \cfrac{\omega}{2\hbar \epsilon_0V} 
\sum\limits_{\breve{\jmath} \breve{k}}
\left[ \sum\limits_{KL} \left\{ \cfrac{\sqrt{n+1} \rho_{K(n+1),L(m-1)}^{\mathrm{eq}}\mu_{kK}^{\breve{\jmath}} \, - \, \sqrt {m-1} \rho_{kn,K(m-2)}^{\mathrm{eq}}\mu_{KL}^{\breve{\jmath}}}{\omega_{kn,L(m-1)} - \omega -i \gamma_{kn,L(m-1)} } \right. 
\mu_{Ll}^{\breve{k}} \sqrt{m}   \right. 
\\
& \qquad& \qquad\qquad -\left. \left. \cfrac{\sqrt{n+2} \rho_{K(n+2),lm}^{\mathrm{eq}}\mu_{LK}^{\breve{\jmath}} \, - \, \sqrt{m} \rho_{L(n+1),K(m-1)}^{\mathrm{eq}}\mu _{Kl}^{\breve{\jmath}}}{\omega_{L(n+1),lm}-\omega-i \gamma_{L(n+1),lm}  } 
\mu_{kL}^{\breve{k}}\sqrt{n+1} \,\right\}e_{\breve{\jmath}}e_{\breve{k}}
\right]
\xi \left(2 \omega \right).
\nonumber
\end{eqnarray}
With these contributions to the density matrix up to second order, we can find and compare the expectation values of the Cartesian components of polarization along direction $\breve{\imath}$, by expressing it as an expansion on the electric field and as a dipole moment:

\begin{equation}
\label{Pesp}
\begin{array}{l}
\left\langle {\hat P}_{\breve{\imath}} \right\rangle  = \epsilon_0 {\rm Tr} \left[ \sum\limits_s \left( \chi_{\breve{\imath}}^{\left( 0 \right)} \lambda ^s \hat\rho ^{\left( s \right)} + \sum\limits_{\breve{\jmath}} \chi_{\breve{\imath}\breve{\jmath}}^{\left( 1 \right)} \lambda ^{s + 1} \hat\rho ^{\left( s \right)} \hat E_{\breve{\jmath}} + \sum\limits_{\breve{\jmath}\breve{k}} \chi _{\breve{\imath} \breve{\jmath} \breve{k}}^{\left( 2 \right)} \lambda ^{s+2} \hat\rho ^{\left( s \right)} \hat E_{\breve{\jmath}} \hat E_{\breve{k}} + ... \right) \right] 
 = N_p {\rm Tr} \left[ \sum\limits_s \lambda ^s \hat\rho ^{\left( s \right)} \hat\mu^{\breve{\imath}} \right].
\end{array}
\end{equation}
Being $N_p$ the number, per unit volume, of active components in the medium. 
\end{widetext}

Unlike the semiclassical theory of susceptibility, this extension requires tracing over the states of the field. Density matrix enters in this trace introducing additional time factors. Susceptibilities of different orders are found by proper matching time oscillating factors for different orders of $\lambda $ in both sides of Eq. (\ref{Pesp}). Note we have used an expansion of polarization in the electric field \cite{Bloembergen} in which the first term is written explicitly. It is usually assumed that such term, the zeroth order susceptibility, either vanishes or can be safely ignored by normalizing polarization with respect to it. The first assumption makes sense in the classical thinking frame: if no electric field is applied, no polarization should be obtained, but from quantum perspective there is always an electric field: vacuum fluctuations cannot be turned off, so it is worth keeping the term explicitly and check if it has relevant effects. The second assumption hides information about spontaneous polarization; this is inconvenient because non-centrosymmetric materials, which are of the highest relevance in nonlinear optics, may present spontaneous polarization. Again, we prefer to keep trace of it explicitly. 

\subsection{Zeroth Order Susceptibility} 
\label{SX0} 
 Comparing terms of zeroth order in $\lambda $ in (\ref{Pesp}), we get: 
 \begin{equation}
 \label{X0}
 \chi_{\breve{i}}^{(0)} = \frac{N_p}{\epsilon_0}\sum\limits_{knl} \rho _{kn,ln}^{\mathrm{eq}}\mu _{lk}^{\breve{i}}.
 \end{equation}
The imaginary part of $ \chi_{\breve{i}}^{(0)} $ is null for every equilibrium state and every value of asymmetry as it is the expectation value of an Hermitian operator. On the other side, its real part will be null only when the following requirements are satisfied simultaneously: 
\begin{enumerate}
 \item[i)] The equilibrium density matrix can be factored as: \begin{equation} 
\label{rhoeq}
\rho _{kn,lm}^{\mathrm{eq}} = \rho_{kl}^{\mathrm{eq}} \rho_{nm}^{\mathrm{eq}},
\end{equation} 	
\item[ii)] The medium reduced density matrix is diagonal: \begin{equation}
\label{condrhoeq}
\rho_{kl}^{\mathrm{eq}} = 0 \quad {\rm for} \quad k \ne l,   
\end{equation}   
\item[iii)]	The diagonal dipole moments of the medium are null: \begin{equation}
\label{condmu}
\mu_{kl} = 0 \quad {\rm for} \quad k = l.
\end{equation}
\end{enumerate}
Conditions (\ref{rhoeq}), (\ref{condrhoeq}) and (\ref{condmu}) are all implicit in the usual semiclassical treatment (see for example ref. \cite{Boyd}). Eq. (\ref{rhoeq}) holds in the case of weak coupling between matter and radiation; if it is met, zeroth order susceptibility is independent of the state of radiation, because the sum over $n$ in (\ref{X0}) equals one. Eq. (\ref{condrhoeq}) holds when thermalization dominates the system dynamics. Both  (\ref{rhoeq}) and (\ref{condrhoeq}) are typically satisfied by open quantum systems in the Born-Markov regime. Eq. (\ref{condmu}) is satisfied by fully symmetric structures. Thus, beyond the Born-Markov regime or when the medium is asymmetric, one would expect zeroth order susceptibility to be not null.

 \subsection{First Order Susceptibility} 
 \label{SX1}
Instead of using a definition, we will calculate $\chi_{\breve{\imath} \breve{\jmath}}^{\left(\omega \right)}$. To do so, we compare terms of first order in $\lambda $ which oscillate in time with frequency $-\omega $, getting (see appendix \ref{App:Appendix}):
\begin{widetext}
\begin{equation}
\label{X1}
\chi_{\breve{\imath} \breve{\jmath}}^{\left(\omega \right)} = \cfrac{1}{\epsilon_0 \hbar \sum\limits_{kn} \rho _{kn,k(n-1)}^{\mathrm{eq}} \sqrt {n} }
 \sum\limits_{knl} \left[  \sum\limits_L  \left\{ \rho_{L(n+1),ln}^{\mathrm{eq}}\mu_{kL}^{\breve{\jmath}}\sqrt{n+1} - \rho_{kn,L(n-1)}^{\mathrm{eq}} \mu_{Ll}^{\breve{\jmath}} \sqrt{n}  \right\} 
\cfrac{ N_p \mu_{lk}^{\breve{\imath}} - \epsilon_0 \chi_{\breve{\imath}}^{\left( 0 \right)} \delta_{lk}}{\omega_{k,l} - \omega - i \gamma_{kn,ln}} \right] .
\end{equation} 
Eq. (\ref{X1}) expresses the linear susceptibility calculated using the second quantization extension, with corrections due to spontaneous susceptibility. Unlike the semiclassical formula, it predicts linear refraction index depends on the state of light and its decay rates. We shall see some applications of this expression in Section III. Note that if all conditions required to make null the zeroth order susceptibility are satisfied, then the decay rates of the media and the radiation will have independent impacts on the dynamics: 
\end{widetext}
	\begin{equation}
	\label{gamma}
	\gamma_{kn,lm} = \gamma_{kl} + \gamma_{nm}.
	\end{equation}
If we further assume radiation rates to be negligible compared to media rates:
	\begin{equation}
	\label{gammarad1}
	\gamma_{nm} << \gamma_{kl}, 
	\end{equation}
then we will arrive to the standard semiclassical expression for linear susceptibility:
\begin{equation}
\label{X1part}
\chi_{\breve{\imath} \breve{\jmath}|\text{sc}}^{\left(\omega \right)} = \cfrac{N_p}{\epsilon_0 \hbar} \sum\limits_{k,l} \cfrac{\left\{ \rho_{ll}^{\mathrm{eq}} - \rho_{kk}^{\mathrm{eq}}  \right\} \mu_{lk}^{\breve{\imath}} \mu_{kl}^{\breve{\jmath}}}
{\omega_{k,l} - \omega - i \gamma_{kl}}  
\end{equation}   
Thus, the standard semiclassical formula for linear susceptibility is a particular case of the one derived here, expected to work well when conditions (\ref{rhoeq}), (\ref{condrhoeq}), (\ref{condmu}) and (\ref{gammarad1}) are satisfied. 

\subsection{Second Order Susceptibility} 
\label{SX2} 
By equating terms of second order in $\lambda$, oscillating with appropriated frequencies in (\ref{Pesp}),  one may find susceptibilities for second order phenomena such as sum and difference frequency generation, optical rectification and SHG. We show here the expression for SGH ($\chi^{\left(2\omega \right)}$): 

\begin{widetext}

\begin{equation}
\label{X2}
\begin{array}{l}
\chi _{\breve{\imath} \breve{\jmath} \breve{k}}^{\left( 2\omega \right)} = 
\cfrac{N_p \epsilon_0^{-1}\hbar^{-2}}{\sum\limits_{k'n'} \rho _{k'n',k'(n'-2)}^{\mathrm{eq}}\sqrt{n'(n'-1)}} \times
\sum\limits_{knl} \left\{  \left(\sum\limits_{k'n'l'} \rho _{k'n',l'n'}^{\mathrm{eq}} \mu_{l'k'}^{\breve{i}}\delta_{l,k}- \mu_{lk}^{\breve{i}} \right) \times \right. 
\\ 
\qquad \quad 
\left( \sum\limits_{KL} \left[ 
\cfrac{\sqrt{n+1} \rho _{K(n+1),L(n-1)}^{\mathrm{eq}} \mu_{kK}^{\breve{\jmath}}-\sqrt{n-1} \rho_{kn,K(n-2)}^{\mathrm{eq}} \mu_{KL}^{\breve{\jmath}}} { \left( \omega_{kl}-2\omega-i\gamma_{kn,ln} \right) \left(\omega_{kL}-i\gamma_{kn,L(n-1)}  \right) } \mu_{Ll}^{\breve{k}}\sqrt{n}
 \right. \right.
\\
\left. \qquad \qquad \qquad\qquad\qquad\qquad \qquad\qquad\qquad  -\cfrac{\sqrt{n+2} \rho _{K(n+2),ln}^{\mathrm{eq}} \mu_{LK}^{\breve{\jmath}}-\sqrt{n} \rho _{L(n+1),K(n-1)}^{\mathrm{eq}} \mu_{Kl}^{\breve{\jmath}}}{\left(\omega_{kl}-2\omega-i\gamma_{kn,ln} \right) \left(\omega_{Ll}-i\gamma_{L(n+1),ln}  \right) } \mu_{kL}^{\breve{k}}\sqrt{n+1}
\right] 
\\ 
\qquad\qquad +
i\cfrac{\sum\limits_{L}  \sqrt{n} \rho_{kn,L(n-1)}^{\mathrm{eq}} \mu_{Ll}^{\breve{\jmath}} -\sqrt{n+1} \rho_{L(n+1),ln}^{\mathrm{eq}} \mu_{kL}^{\breve{\jmath}} } 
{ \left( \omega_{kl}-\omega-i\gamma_{kn,ln} \right)  \sum\limits_{k'n'} \sqrt{n'} \rho_{k'n',k'(n'-1)}^{\mathrm{eq}} } 
\\
 \left. \left. \qquad\qquad\qquad\qquad
\times

 \sum\limits_{k'n'} \left[ \cfrac{\sqrt{n'}}{\gamma_{kn,k(n-1)}}\sum\limits_{L'} \left\{ \sqrt{n'-1} \rho_{k'n',L'(n'-2)}^{\mathrm{eq}} \mu_{L'k'}^{\breve{k}} - \sqrt{n'+1} \rho_{L'(n' +1),k'(n'-1)}^{\mathrm{eq}} \mu_{k'L'}^{\breve{k}} \right\}   \right]  
	
\right)
 \right\} .
\end{array} 
\end{equation}  
The second order nonlinear susceptibility calculated this way is also radiation state dependant. 

One would expect that imposing in (\ref{X2}) all conditions used to obtain the semiclassical expression for   $\chi_{\breve{i}}^{(\omega)}$ --(\ref{rhoeq}), (\ref{condrhoeq}), (\ref{condmu}) and (\ref{gammarad1})--, would lead to the semiclassical standard  expression for $\chi_{\breve{i}}^{(2\omega)}$; interestingly, one achieves a very similar one, but not exactly the same:

\begin{equation}
\label{X2class}
\begin{array}{l}
\chi _{\breve{\imath} \breve{\jmath} \breve{k}|\text{sc}}^{\left( 2\omega \right)} = 
N_p \epsilon_0^{-1}\hbar^{-2} 
\sum\limits_{klL} \left\{ \cfrac{1}{\omega_{lk}-2\omega-i\gamma_{lk} } 
\left(  \cfrac{(\rho_{kk}^{\mathrm{eq}} - \rho_{LL}^{\mathrm{eq}} )\mu_{kl}^{\breve{i}} \mu_{lL}^{\breve{\jmath}} \mu_{Lk}^{\breve{k}} }{  \omega_{Lk}-i\gamma_{Lk}} -
\cfrac{(\rho _{LL}^{\mathrm{eq}} - \rho_{ll}^{\mathrm{eq}})\mu_{kl}^{\breve{i}} \mu_{Lk}^{\breve{\jmath}} \mu_{lL}^{\breve{k}}} {\omega_{lL}-i\gamma_{lL}} 
\right)
\right\} .
\end{array} 
\end{equation}  
This expression should be compared to the standard semiclassical formula before intrinsic permutation symmetry is imposed. The resonance $\omega_{zz'} - i\gamma_{zz'} $ of the terms inside the parenthesis in (\ref{X2class}) take the form $\omega_{zz'} +\omega - i\gamma_{zz'} $ in the standard formula \footnote{See Eq. (3.6.13) in pag. 172 of \cite{Boyd}. A change of dummy indexes was made in (\ref{X2class}) to ease comparison.}.  The frequency $\omega$ cancels out in our procedure as a consecuence of the contribution of the radiation Hamiltonian. Eq. (\ref{X2class}) indicates that even in a semiclassical scenario, the standard formula is a less precise approximation than the second quantization expression. This difference could be used to test the validity of the present work.       
\end{widetext}

\section{APLICATION TO QUANTUMS DOTS} 
\label{SAQD} 
This theory can be applied to many kinds of quantum systems; as an example, we shall consider a QD with two active levels $ \ket{1} =\left( 1\;\; 0 \right)^T $ and $\ket{0} =\left( 0\;\; 1 \right)^T $, having energies $\varepsilon _1=\hbar \omega_x $ y $\varepsilon _0 = 0$ respectively. The QD is assumed to be inside a cavity and interacting resonantly with a confined mode  of frequency $\omega$ in a Coherent, Thermal or Fock state of radiation. We define the detuning as the difference between the frequency of the field and the frequency of the QD: $\Delta=\omega -\omega_x$. The state of the complete system is assumed to fulfill condition (\ref{rhoeq}) and the states of the QD to be described (in the $ \{\ket{1},\ket{0} \}$ basis), by means of the density matrix:
\begin{equation}
\label{rhoeqmat}
\hat \rho_{\mathrm{QD}}^{\mathrm{eq}} = \left(\begin{array}{cc}
\alpha &\beta \\
\beta^* & 1 - \alpha
\end{array} \right),
\end{equation}
with $\alpha \in \left[ 0,1 \right]$ and $\beta \in \mathbb{C}$. In particular, we will investigate the following QD states: Pure Excited (PE) State: $ \hat \rho_{\mathrm{QD}}^{\mathrm{PE}} = \ket{1} \bra{1} $; Pure Ground (PG) State: $ \hat \rho_{\mathrm{QD}}^{\mathrm{PG}} = \ket{0} \bra{0} $; Maximally Mixed (MM) State: $ \hat \rho_{\mathrm{QD}}^{\mathrm{MM}} = \frac{1}{2}  \ket{1} \bra{1} + \frac{1}{2}  \ket{0} \bra{0} $ and Maximal Superposition (MS) State: $ \rho_{\mathrm{QD}}^{\mathrm{MS}} = \frac{1}{2} \left(\ket{1} + \ket{0}\right) \left(\bra{1} +\bra{0} \right)$.  
             
In what follows, we assume the decay rates obey Eq. (\ref{gamma}), with the QD rates given by: 
\begin{equation}
\label{QD}
\begin{matrix}
\gamma _{kk}= \gamma,
\\
\; \; \gamma _{kl}= \gamma_{\mathrm{d}}; \quad  (l \neq k),
\end{matrix}
\end{equation}
where, $\gamma$ is the spontaneous emission rate and $\gamma_{\mathrm{d}}$ is the dephasing rate of the QD. 
\\For the field, we shall explore different scenarios: 
\begin{numcases}{\gamma _{nm} =}
0; \label{gammarad:a}\\ 
\kappa; \label{gammarad:b} \\ 
\kappa \left( m + n - 0.2 \right). \label{gammarad:c}
\end{numcases}
Here a) corresponds to the semiclassical scenario, in which radiation states have no effect on the dynamics; b) to a situation in which all radiation states have the same effect, and c) to a situation in which decay from excited levels to the ground level and pumping from the ground level to higher states, influence the dynamics.
        
We assume, besides, the dipole moment to have equal components along all cartesian directions: $ \hat {\boldsymbol\mu}= \hat{\mu}(\mathbf{e}_1 + \mathbf{e}_2 + \mathbf{e}_3) $. Following Savenko et al. \cite{Savenko}, we model an asymmetric nonferroelectric QD by the means of the dipole operator:
\begin{eqnarray}
{\bf \hat{\mu}^{\breve{\imath}}}=
\left( \begin{matrix}
\mathbf{\mu}_{ee} & \mathbf{\mu}_{eg} \\
\mathbf{\mu}_{ge} & \mathbf{\mu}_{gg}  \\
\end{matrix} \right) &=&
\cfrac{\mathbf{\mu}_{ee}}{2} \left( \mathbb{I}+\mathbf{\hat{\sigma}}_z \right) + 
\cfrac{\mathbf{\mu }_{gg}}{2} \left( \mathbb{I}-\mathbf{\hat{\sigma}}_z \right) \qquad
\\
&+& \mathbf{\mu }_{eg} \mathbf{\hat{\sigma}}^{+}+\mathbf{\mu}_{ge} \mathbf{\hat{\sigma}}^- ,
\nonumber
\end{eqnarray}
Where we have used the Pauli matrices: $ \hat{\sigma}_x \pm  i \hat{\sigma}_y = \hat{\sigma}^{\pm}, \hat{\sigma}_z $. Here, asymmetry information is carried by diagonal matrix elements, allowed to take nonzero values. For nonferroelectric QDs, $\mu_{gg}$ is proportional to the size of the elementary cell of the crystal lattice and $\mu_{ee}$ is proportional to the size of the QD; therefore, we assume $\mu_{ee} > \mu_{gg} $ and nondiagonal matrix elements to be real: 
\begin{equation}
\label{mu}
{\hat\mu}^{\breve{\imath}} = g\sqrt {2 \epsilon_0 \hbar V \omega_x^{-1}} \left( {\begin{array}{*{2}{c}} 0.1A&1\\1&{A} \end{array}} \right).
\end{equation}
Here, $g$ is the coherent coupling strength (Rabi) between the QD and the cavity mode and $ A \in \left[ 0,1 \right]$ quantifies the asymmetry: $A=0$ for fully symmetric QDs and $A=1$ for highly asymmetric QDs.   

Finally, for all graphs shown below, we use the experimental values: $\kappa  = 2\pi  \times 27 \, \rm{GHz}$, $g = 2\pi \times 25 \: \, \rm{GHz}$,  $\gamma = 2\pi \times 100 \, \rm{GHz}$ and $\gamma_d = 2\pi \times 10 \, \rm{GHz}$, estimated in \cite{Majumdar, Majumdar_thesis} for a QD confined in a nanocavity.
  
\subsection{Spontaneous Susceptibility of a QD:}
\label{Spontaneous Susceptibility of a QD:}
\begin{figure}[tb]
	\includegraphics[width=0.48\textwidth]{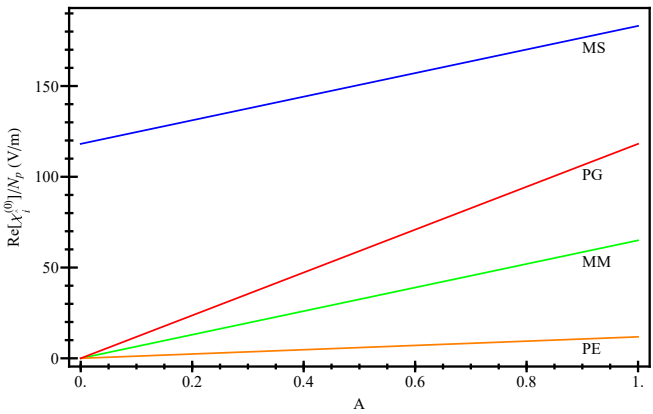}
	\caption{  $\operatorname{Re}[\chi_{_{\breve{\imath}}}^{(0)}] \neq 0 $ for  the MS State even if the QD, is fully symmetric $(A=0)$. All figures in this paper assume: separability condition, (\ref{rhoeq}), is valid; $V=5\times10^{-2} \mu m^{3}, \; \omega_x =2.05 \; \textrm{THz}, \; \epsilon_0=1. $ All other parameters take the values already mentioned in the text. Please note that colored figures are available only in the on-line version of this paper.}
	\label{X0vsA}
\end{figure}

The real part of zeroth order susceptibility for all equilibrium states mentioned above is shown in Fig. \ref{X0vsA}. Condition (\ref{condrhoeq}) is satisfied by all those states except the MS, therefore, only that state has a not null $\chi _{\breve{\imath}}^{(0)}$ when (\ref{condmu}) is satisfied (i.e. when $A=0$). 
 The imaginary part of $ \chi_{_{\breve{\imath}}}^{(0)}$, not shown, is null for every equilibrium state and every value of asymmetry.

We thus may interpret  $\chi _{\breve{\imath}}^{(0)}$ as a polarization the QD has spontaneously, due to its asymmetry, or may acquire by reaching the MS State. This is consistent with the findings of another extension of semiclassical theory of susceptibility \cite{SONOP}.

\subsection{Linear optical response of a QD interacting with coherent light:}
\label{Linear optical response of a QD:}  
\begin{figure}[bt]
	\includegraphics[width=0.48\textwidth]{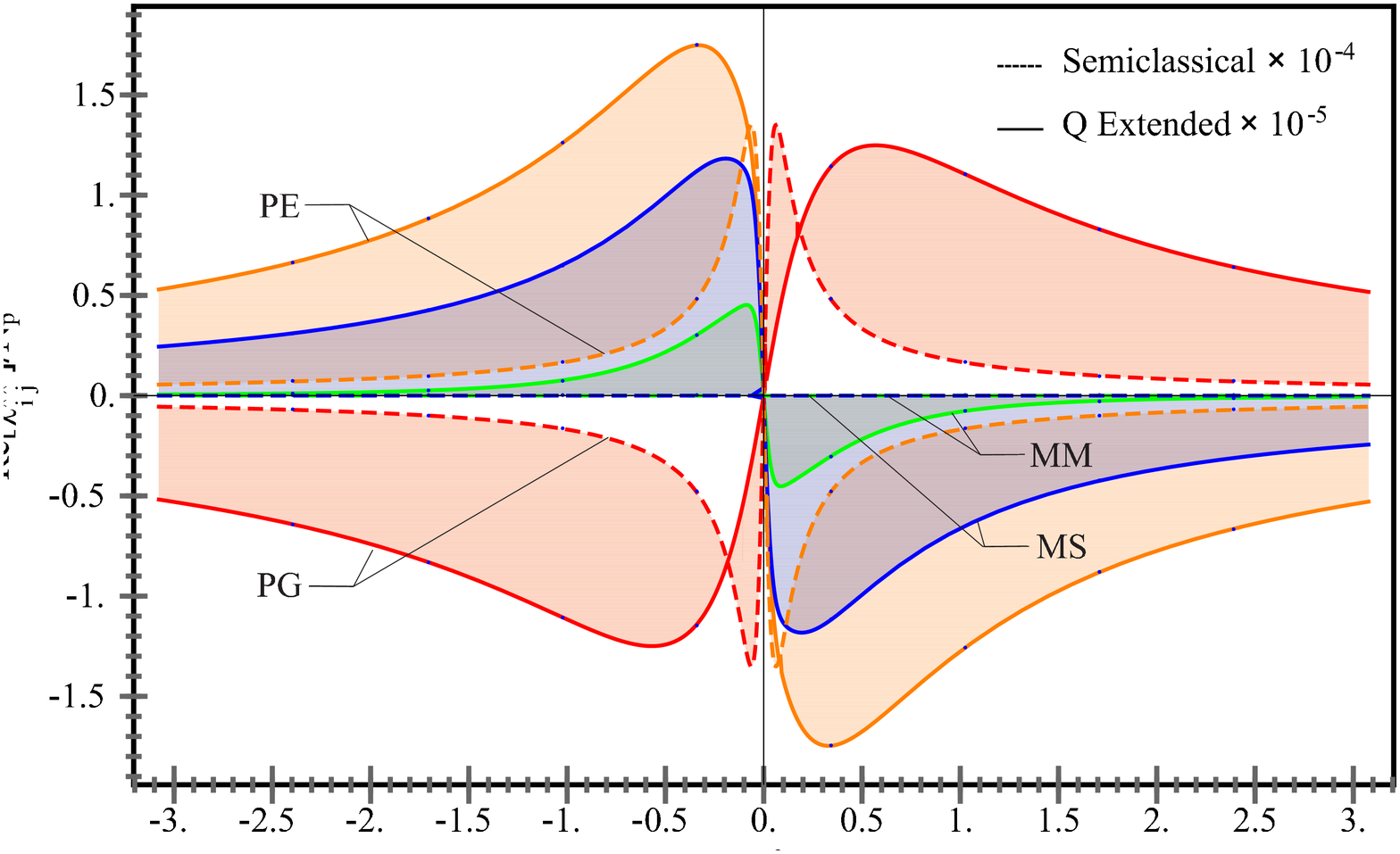}
	\caption{The real part of the linear susceptibility of an asymmetric ($A=1$) QD as a function of detuning, when it is interacting with a Coherent State of light with average photon number $\bar{n}=3$. Semiclassical predictions $-$dotted lines, decay rates of the type (\ref{gammarad:a}$-$ differ from quantum extension predictions $-$continuous lines, decay rates of the type (\ref{gammarad:c})$-$ in the widths and values of resonances. The quantum extension predicts nonzero values for susceptibilities of QDs in equal population equilibrium states, while the semiclassical theory predicts they should be null $-$see Eq. (\ref {X1part})$-$. Decay rates of the types (\ref{gammarad:a}) and (\ref{gammarad:c}) for the semiclassical and quantum regimes respectively, shall be used from now on.}
	\label{X1vsWClass-Q}
\end{figure}

Inserting the matrix elements of a coherent field of average photon number $\bar{n}$ in Eq. (\ref{X1}), leads to the following expression for susceptibility:
\begin{eqnarray}
\label{X1coherente}
\chi_{\breve{\imath} \breve{\jmath}}^{(\omega)} &=&  \cfrac{e^{-\bar{n}}}{\epsilon_0 \hbar}\sum\limits_{knl} \left[ \sum\limits_L \left\{ \rho_{Ll}^{\mathrm{eq}}\mu_{kL}^{\breve{\jmath}}  \cfrac{\bar{n}^{n}}{n!} \right. \right.  
\\ 
&\qquad&  - \left. \left. \rho_{kL}^{\mathrm{eq}}\mu_{Ll}^{\breve{\jmath}} \cfrac{\bar{n}^{(n-1)}}{(n-1)!}  \right\} 
\cfrac{N_p \mu_{lk}^{\breve{\imath}} - \epsilon_0 \chi _{\breve{\imath}}^{(0)}\delta_{lk}} {\omega_{kl} - \omega - i\gamma_{kn,ln}} \right] . 
\nonumber 
\end{eqnarray}
The values predicted by (\ref{X1coherente}) are shown in Fig. \ref{X1vsWClass-Q} $-$continuous lines$-$, compared to those predicted by the semiclassical theory $-$Eq. (\ref{X1part}), dotted lines $-$. Both treatments agree in attributing a resonant nature to linear susceptibility, but they differ in the widths and heights of resonances, which are affected by the radiation states and their decay rates in the quantum extension. In this particular situation, the values predicted by the semiclassical theory and the quantum extension differ radically, one of the most remarkable differences appears for equal population equilibrium states of the matter subsystem; as we show somewhere else \cite{SONOP}, that is a consequence of forbidding the MS as a possible equilibrium state in the semiclassical theory. 
     
\begin{figure}[tb]
	\includegraphics[width=0.48\textwidth]{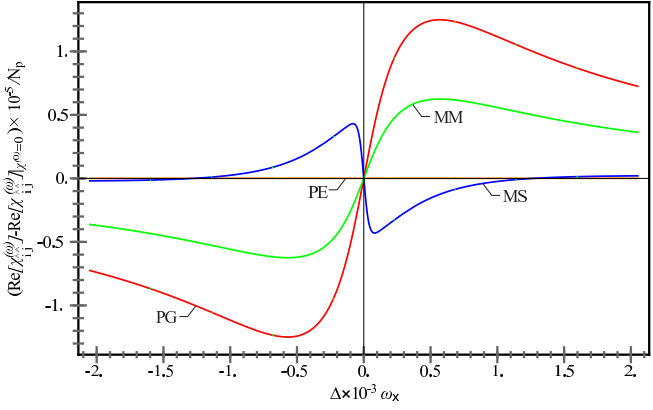}
	\caption{Differential contribution of zeroth order susceptibility to the real part of linear susceptibility of an asymmetric QD $(A=1)$ interacting with coherent radiation ($\bar{n}=3$), calculated near the $\Delta=0$ resonance.}
	\label{ContrX0}
\end{figure}

According to the present extension, $\chi^{(0)}$ may significantly contribute to linear susceptibilty near the resonance, as shown in Fig. \ref{ContrX0}, where we plot the difference between the linear susceptibility as given by Eq. (\ref{X1coherente}) with and without the zeroth order contribution. This is consistent with that said in sec. (\ref{SX1}): semiclassical approximation can only work well when all conditions for $ \chi^{(0)}=0$ are met. \\

\begin{figure}[tb]
\includegraphics[width=0.48\textwidth]{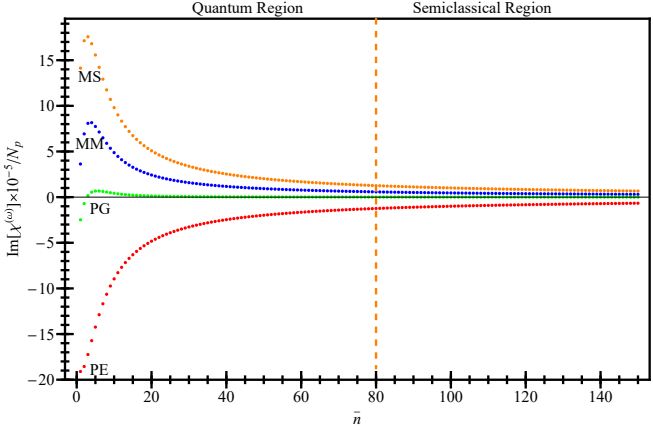}
\caption{The imaginary part of the linear susceptibility of a QD interacting with coherent light, shown for different equilibrium states of the QD, as a function of average photon number.  A detuning of $2 \times 10^{-4} \omega_x$ was assumed. Variations of absorption drop by a factor of $10^2$ when reaching $\overline{n}=80$.} 
\label{ImX1vsn}
\end{figure}

The dependence of $\chi_{\breve{\imath} \breve{\jmath}}^{(\omega)}$ on radiation states is reflected, for coherent states, in a variation of susceptibility, and thus of refraction index, with average photon number, as can be seen in Fig. \ref{ImX1vsn}. 

Unlike the intensity dependent refraction index, which is a nonlinear third order phenomenon typically requiring very high intensities, the index variation predicted here is a linear phenomenon, more appreciable when the average photon number is small. In fact, as photon number increases, susceptibility becomes less sensible to it $-$Fig. \ref{ImX1vsn}. One may speak of a limit for reaching a semiclassical behavior of susceptibility in the sense it will almost be independent of $\overline{n}$. In Fig. \ref{ImX1vsn} we assume that limit to be found when $\frac{d}{d \overline{n}} \left( \ \chi_{\breve{\imath} \breve{\jmath}}^{(\omega)}\right)  < 10^{-2}$. Thus, the ``semiclassical behavior" is reached for $ \overline{n} = 80 $. In general, we hope the semiclassical approximation, with a suitable correction on a global decay rate, to work well for interactions with most common coherent light sources, characterized by an average photon number above several thousands. On the other side, this result is interesting as it suggests that even a Poissonian distribution of light of coherent origin cannot be properly described by the semiclassical approximation if the average photon number is low.

\subsection{Linear optical response of a QD interacting with diagonal states of radiation}
This extension does not allow to calculate exact expressions for susceptibility if interaction takes place with diagonal states of light, i.e. Fock or Thermal States, because when condition (\ref{rhoeq}) is valid, the matrix elements of the field take the form: $\rho^{\mathrm{eq}}_{mn} = \delta_{mN}\delta_{nN}$ and this make the fraction (\ref{X1}) undetermined. Nonetheless, we can use auxiliary radiation states to calculate approximate expressions. We can assume, for example, the QD approaches to a Number State passing through the state:	
\begin{eqnarray}
\label{rhoarb}
\rho_{mn}^{\mathrm{aux}} &=& \delta_{mn} \delta_{nN}
\\
&+& e^{ - \varphi } \left( \delta_{m,N - 1} \delta _{n,N + 1} + \delta_{m,N + 1} \delta_{n,N - 1} \right), \nonumber
\end{eqnarray}
which converges to a Number State when $\varphi \to \infty $. Replacing (\ref{rhoarb}) in (\ref{X1}) and taking the limit we find: 
\begin{eqnarray}
\label{X1number}
\chi_{\breve{\imath} \breve{\jmath}}^{(\omega)} &=&   \cfrac{1}{\epsilon_0 \hbar} \sum\limits_{k,l,L} \left[ \left( \cfrac{\rho_{Ll}^{\mathrm{eq}} \mu_{kL}^{\breve{\jmath}}}{ \omega_{kl} - \omega - i \gamma_{kN,lN}} \right. \right.  
\\
&-&\left. \left. \cfrac{\rho_{kL}^{\mathrm{eq}} \mu_{Ll}^{\breve{\jmath}}}{ \omega_{kl} - \omega -i \gamma_{k(N+1),l(N+1)} } \right) 
\left(N_p \mu_{lk}^{\breve{\imath}}- \epsilon_0 \chi _{\breve{\imath}}^{(0)} \delta_{lk} \right) \right]. \nonumber
\end{eqnarray}
Where fulfillment of conditions (\ref{gamma}) and (\ref{gammarad:a}) was assumed.

\begin{figure}[tb]
	\includegraphics[width=0.48\textwidth]{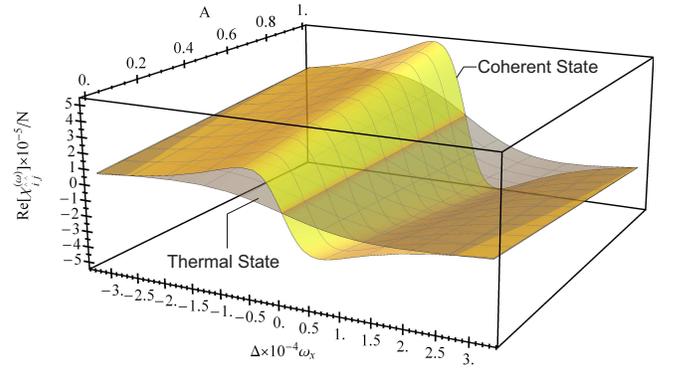}
	\caption{The real part of the linear susceptibility for a QD in the MS equilibrium state, interacting with a Coherent State with average photon number $\bar{n}=1$ or with a Fock State with $n=1$.}
	\label{X1vsDvsA}
\end{figure}
Fig. \ref{X1vsDvsA} compares the real parts of the linear susceptibility of a QD, when interacting with a Number State, and a Coherent State. It is apparent there that, near the resonance, different states of radiation experience different refraction indexes, regardless of the value of asymmetry.
\\
By using a similar procedure we find Thermal States to have the same susceptibility as Coherent States (\ref{X1}). Susceptibilities calculated this way will depend on the path followed to reach the equilibrium state. Inconvenient as this is, it may be more accurate than using the semiclassical expression, which has been proved to be a particular case of Eq. (\ref{X1}).

\subsection{SHG in a QD:} 
\label{SHG in a QD:} 
\begin{figure}[tb]
\includegraphics[width=0.48\textwidth]{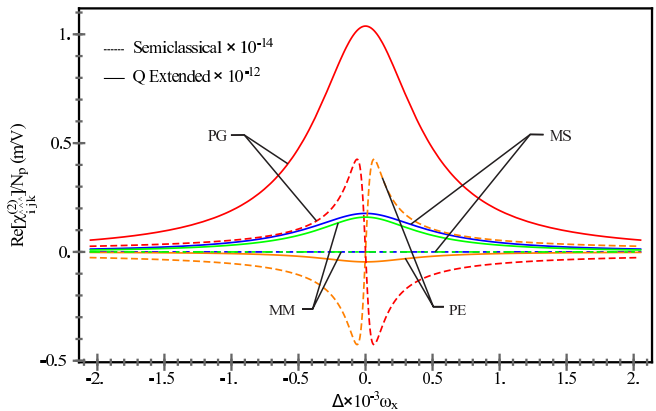}
\caption{Comparison of the semiclassical $-$dotted lines$-$ and the quantum extension predictions $-$continuous lines$-$, for the real part of $\chi^{(2\omega)}$ of an asymmetric ($A=1$) QD as a function of detuning. In all the equilibrium states considered for the quantum extension, the QD was assumed to interact with coherent light of average photon number $\bar{n}=3$.} 
\label{X2vsWClass-Q}
\end{figure}

Second order susceptibility values, predicted by means of the quantum extension, may present dramatic differences with those predicted by the semiclassical theory as can be seen in Fig. \ref{X2vsWClass-Q} for the case of a QD interacting with a very low intensity Coherent State of radiation. Note the quantum extension predicts susceptibility values two orders of magnitude larger than the semiclassical theory.     

\begin{figure}[tb]
	\includegraphics[width=0.48\textwidth]{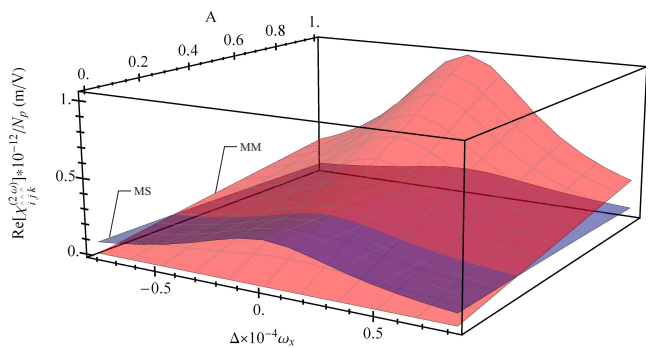}
	\caption{The real part of second order susceptibility, for the case of SHG, of a QD in a MM (MS) State interacting with a Coherent State of light ($\bar{n}=1$), as a function of detuning and asymmetry. Observe that, even if the QD is fully symmetric ($A=0$) the value of $\chi^{(2\omega)}$ may be not null if we allow the MS to be one of its possible equilibrium states.}
	\label{X2vsDvsA}
\end{figure}

Fig. \ref{X2vsDvsA} shows the second order susceptibility for two different equilibrium states of a QD interacting with a Coherent State of radiation with $\bar{n}=1$, as a function of detuning and asymmetry. Note $\chi^{(2\omega)} \neq 0$ if $\chi^{(0)} \neq 0$ and this last one may take nonzero values even for null asymmetry, if the active medium is in a MS State. Thus, $\chi^{2\omega)}$ may have nonzero values for a centrosymmetric QD in such a state (see also Fig. \ref{X2vsWClass-Q}). Again, this is consistent with the findings of a previous extension of semiclassical theory of susceptibility \cite{SONOP}.     

\begin{figure}[tb]
	\includegraphics[width=0.48\textwidth]{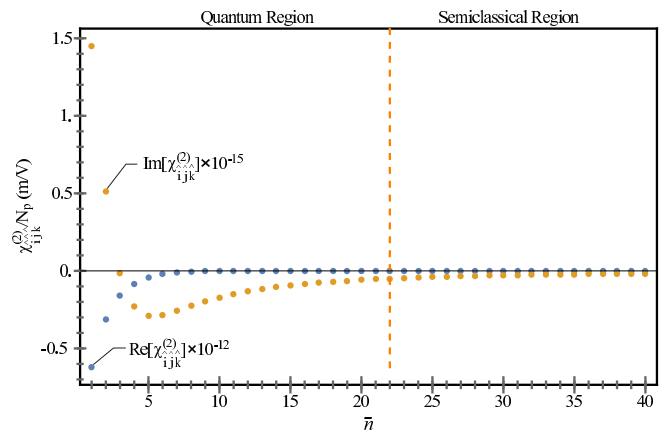}	
	\caption{Real and imaginary parts of second order susceptibility, as a function of average photon number, for a QD in the MS state, interacting with a Coherent State of light.}
	\label{X2vsn}
\end{figure}

Second order susceptibility is also photon number dependent, as can be seen in Fig. \ref {X2vsn}. Again, we see the statistical distribution of light and its origin to be not sufficient when it comes to label a source as a classical one.
 
\section{DISCUSSION AND CONCLUSIONS} 
\label{SDC} 
We have extended the semiclassical theory of electrical susceptibility to one in which the electric field is quantized. The expressions derived predict new phenomena such as state and photon number dependent susceptibilities both in the linear and nonlinear regime. This implies that the refraction index is a function of intensity not only for strong electromagnetic fields, as semiclassical nonlinear optics predicts, but for very weak fields too, as shown in Fig . This may have impact, for instance, in the focal length of lenses used in experiments with ultralow intensity light and in calibration of photon counters, as the absorption coefficients may be affected by the average number of incident photons. Generally speaking, susceptibility should not be considered a property of the media, but of its interaction with radiation.
\\
We have found the semiclassical formula of linear susceptibility to be a particular case of that derived here, expected to be inadequate if quantum aspects of light are conspicuous, as when one deals with Fock States or Thermal or Coherent States in ultra low intensity regime. In fact, the semiclassical approximation should work well only when conditions (\ref{rhoeq}), (\ref{condrhoeq}), (\ref{condmu}) and (\ref{gammarad1}) are satisfied. \\
On the other side, we have found that even in the semiclassical limit, the second order susceptiblity expression should have a correction in the positions and widths of the resonances, when compared to the standard semiclassical formula. Besides, it is to be expected that $\chi^{(2\omega)}$ should be greater at least two orders of magnitude if radiation is confined in a cavity; this opens the possibility to explore for higher efficiency SHG devices. 

The validity of this extension can be tested experimentally through its predictions. Let us briefly mention two examples of such experiments: a) measuring the light transmitted through a microcavity with embedded QDs should reveal a dependence of absorption coefficient on intensity, when average number of incident photons is of the order of tens; b) measuring the transmitted intensity of light of a certain frequency in a Fabry-Pérot cavity with an active medium, should reveal differences between a thermal (filtered) source and a coherent source, due to the different refraction indexes experienced by each state. Confirming this will show, besides, that a Poissonian statistics and a coherent distribution are not enough to guarantee a classical behavior of light.   

We are currently working on Lindblad Born-Markov master equation, to take into account, in a more realistic way, the complex non-Hamiltonian processes in the framework of a theory of open quantum systems. 

\section{ACKNOWLEDGMENTS}  
J.D.G. acknowledges support from Dinntec SAS. B.A.R. and J.D.G. acknowledge support from the Comité para el Desarrollo de la Investigación (CODI) of the Universidad de Antioquia through: "Proyecto: Caracterización Experimental de Fotones Individuales" and "Proyecto de Sostenibilidad, Grupo de Física Atómica".

\appendix  \section{PROCEDURE FOR FINDING THE LINEAR SUSCEPTIBILITY \label{App:Appendix}}  

Equating terms of first order in $\lambda$ in (\ref{Pesp}) we get:
\begin{equation}
\label{Pesp_1}
\epsilon_0 \text{Tr}\left[ \chi_{\breve{\imath}}^{(0)} \rho^{(1)} + \sum\limits_{\breve{\imath}} \chi_{\breve{\imath}\breve{\jmath}}^{(1)} \rho^{(0)} E_{\breve{\jmath}} \right] = N \text{Tr} \left[ \rho^{(1)} \mu^{\breve{\imath}} \right],
\end{equation}	

which, using the completeness of the bases of states for field and matter we may rewrite as:
\begin{eqnarray}
\label{toX1}
&\epsilon_0& \sum\limits_{knlm} \sum\limits_{\breve{\jmath}}\chi _{\breve{\imath}\breve{\jmath}}^{(1)}\bra{kn}\rho^{(0)}\ket{lm} \bra{lm} \hat{E}^{\breve{\jmath}} \ket{kn} \qquad\qquad\qquad \nonumber
 \\ 
&=& \sum\limits_{kn} \left( \sum\limits_{lm} N \bra{kn} \hat{\rho}^{(1)} \ket{lm} \bra{lm} \mu^{\breve{\imath}} \ket{kn} \right.   \\ 
&\qquad&  \left. \qquad\qquad\qquad\qquad\qquad -\cfrac{}{}\epsilon_0\chi _{\breve{\imath}}^{(0)} \bra{kn} \hat{\rho}^{(1)} \ket{kn} \right).  \nonumber
\end{eqnarray}
Note $ \rho^{(1)} $ does not represents the full density matrix and thus the condition $\sum\limits_{kn}{\rho _{knkn}^{(1)} = 1} $ needs not to be satisfied. 
\\
Replacing the explicit form of electric field in Eq. (\ref{toX1}) we achieve:
\begin{equation}
	\label{toX1_2}
\begin{matrix}
\sum\limits_{knm \breve{\jmath}} \chi_{\breve{\imath}\breve{\jmath}}^{(1)} e^{\breve{\jmath}} \rho_{knkm}^{(0)} \left\langle  m \right|\left( a e^{-i\omega t}- a^{\dagger} e^{i\omega t} \right) \left|n\right\rangle \qquad 
\\
\quad \quad = i \left(\cfrac{\hbar \omega \epsilon_{0}}{2 V} \right)^{- \frac{1}{2}} \sum\limits_{knl} \rho_{knln}^{(1)} \left(\epsilon_{0} \chi _{\breve{\imath}}^{(0)} \delta_{lk} -N\mu_{lk}^{\breve{\imath}} \right). 
\end{matrix}
\end{equation}

Given the temporal dependence of first order density matrix, equation (\ref{toX1_2}) imposes one condition for terms oscillating with frequency $\omega$ and other for terms oscillating with frequency $-\omega$. For positive frequencies we have:	
\begin{equation}
	\label{Pesp_1 desp w}
\begin{matrix}
\sum\limits_{knm \breve{\jmath}} \chi_{\breve{\imath}\breve{\jmath}}^{(\omega)}e^{\breve{\jmath}} \rho_{knkm}^{(0)} \sqrt{n} \delta_{m,n-1}  \qquad \qquad\qquad\qquad \qquad
\\
\quad= i \left(\cfrac{\hbar \omega \varepsilon_{0}}{2 V} \right)^{- \frac{1}{2}} \sum\limits_{knl \breve{\jmath}} \rho _{knln}^{\left(\omega_{\breve{\jmath}} \right)} e^{\breve{\jmath}} \left( \epsilon_{0} \chi_{\breve{\imath}}^{(0)} \delta_{lk} - N\mu_{lk}^{\breve{\imath}}\right).
\end{matrix}
\end{equation}	

To deduce the components of the tensor $ \boldsymbol{\chi }^{(\omega)}$ from this point, we only need to compare expressions (\ref{Pesp_1 desp w})  and (\ref{rhowf}). Following a similar procedure we can calculate the susceptibility for any other frequency.

\bibliography{All-QT-Suscept}

\begin{thebibliography}{24}%
\makeatletter
\providecommand \@ifxundefined [1]{%
 \@ifx{#1\undefined}
}%
\providecommand \@ifnum [1]{%
 \ifnum #1\expandafter \@firstoftwo
 \else \expandafter \@secondoftwo
 \fi
}%
\providecommand \@ifx [1]{%
 \ifx #1\expandafter \@firstoftwo
 \else \expandafter \@secondoftwo
 \fi
}%
\providecommand \natexlab [1]{#1}%
\providecommand \enquote  [1]{``#1''}%
\providecommand \bibnamefont  [1]{#1}%
\providecommand \bibfnamefont [1]{#1}%
\providecommand \citenamefont [1]{#1}%
\providecommand \href@noop [0]{\@secondoftwo}%
\providecommand \href [0]{\begingroup \@sanitize@url \@href}%
\providecommand \@href[1]{\@@startlink{#1}\@@href}%
\providecommand \@@href[1]{\endgroup#1\@@endlink}%
\providecommand \@sanitize@url [0]{\catcode `\\12\catcode `\$12\catcode
  `\&12\catcode `\#12\catcode `\^12\catcode `\_12\catcode `\%12\relax}%
\providecommand \@@startlink[1]{}%
\providecommand \@@endlink[0]{}%
\providecommand \url  [0]{\begingroup\@sanitize@url \@url }%
\providecommand \@url [1]{\endgroup\@href {#1}{\urlprefix }}%
\providecommand \urlprefix  [0]{URL }%
\providecommand \Eprint [0]{\href }%
\providecommand \doibase [0]{http://dx.doi.org/}%
\providecommand \selectlanguage [0]{\@gobble}%
\providecommand \bibinfo  [0]{\@secondoftwo}%
\providecommand \bibfield  [0]{\@secondoftwo}%
\providecommand \translation [1]{[#1]}%
\providecommand \BibitemOpen [0]{}%
\providecommand \bibitemStop [0]{}%
\providecommand \bibitemNoStop [0]{.\EOS\space}%
\providecommand \EOS [0]{\spacefactor3000\relax}%
\providecommand \BibitemShut  [1]{\csname bibitem#1\endcsname}%
\let\auto@bib@innerbib\@empty
\bibitem [{\citenamefont {Grynberg}\ \emph {et~al.}(2010)\citenamefont
  {Grynberg}, \citenamefont {Aspect},\ and\ \citenamefont {Fabre}}]{Aspect}%
  \BibitemOpen
  \bibfield  {author} {\bibinfo {author} {\bibfnamefont {G.}~\bibnamefont
  {Grynberg}}, \bibinfo {author} {\bibfnamefont {A.}~\bibnamefont {Aspect}}, \
  and\ \bibinfo {author} {\bibfnamefont {C.}~\bibnamefont {Fabre}},\
  }\href@noop {} {\emph {\bibinfo {title} {Introduction to Quantum Optics From
  the Semi-classical Approach to Quantized Light}}}\ (\bibinfo  {publisher}
  {Cambridge University Press},\ \bibinfo {year} {2010})\BibitemShut {NoStop}%
\bibitem [{\citenamefont {Narozhny}\ \emph {et~al.}(1981)\citenamefont
  {Narozhny}, \citenamefont {Sanchez-Mondragon},\ and\ \citenamefont
  {Eberly}}]{Eberly}%
  \BibitemOpen
  \bibfield  {author} {\bibinfo {author} {\bibfnamefont {N.~B.}\ \bibnamefont
  {Narozhny}}, \bibinfo {author} {\bibfnamefont {J.~J.}\ \bibnamefont
  {Sanchez-Mondragon}}, \ and\ \bibinfo {author} {\bibfnamefont {J.~H.}\
  \bibnamefont {Eberly}},\ }\href {\doibase 10.1103/PhysRevA.23.236} {\bibfield
   {journal} {\bibinfo  {journal} {Phys. Rev. A}\ }\textbf {\bibinfo {volume}
  {23}},\ \bibinfo {pages} {236} (\bibinfo {year} {1981})}\BibitemShut
  {NoStop}%
\bibitem [{\citenamefont {Burnham}\ and\ \citenamefont
  {Weinberg}(1970)}]{Burnham}%
  \BibitemOpen
  \bibfield  {author} {\bibinfo {author} {\bibfnamefont {D.~C.}\ \bibnamefont
  {Burnham}}\ and\ \bibinfo {author} {\bibfnamefont {D.~L.}\ \bibnamefont
  {Weinberg}},\ }\href@noop {} {\bibfield  {journal} {\bibinfo  {journal}
  {Phys. Rev. Lett.}\ }\textbf {\bibinfo {volume} {25}},\ \bibinfo {pages} {84}
  (\bibinfo {year} {1970})}\BibitemShut {NoStop}%
\bibitem [{\citenamefont {Shih}(2003)}]{Shih}%
  \BibitemOpen
  \bibfield  {author} {\bibinfo {author} {\bibfnamefont {Y.}~\bibnamefont
  {Shih}},\ }\href {http://stacks.iop.org/0034-4885/66/i=6/a=203} {\bibfield
  {journal} {\bibinfo  {journal} {Rep. Prog. Phys.}\ }\textbf {\bibinfo
  {volume} {66}},\ \bibinfo {pages} {1009} (\bibinfo {year}
  {2003})}\BibitemShut {NoStop}%
\bibitem [{\citenamefont {Boyd}(2008)}]{Boyd}%
  \BibitemOpen
  \bibfield  {author} {\bibinfo {author} {\bibfnamefont {R.~W.}\ \bibnamefont
  {Boyd}},\ }\href@noop {} {\emph {\bibinfo {title} {Nonlinear Optics, Third
  Edition}}}\ (\bibinfo  {publisher} {Academic Press},\ \bibinfo {year}
  {2008})\BibitemShut {NoStop}%
\bibitem [{\citenamefont {Shen}(1984)}]{Shen}%
  \BibitemOpen
  \bibfield  {author} {\bibinfo {author} {\bibfnamefont {Y.~R.}\ \bibnamefont
  {Shen}},\ }\href@noop {} {\emph {\bibinfo {title} {Principles of Nonlinear
  Optics}}}\ (\bibinfo  {publisher} {Wiley},\ \bibinfo {year}
  {1984})\BibitemShut {NoStop}%
\bibitem [{\citenamefont {Sugawara}\ \emph {et~al.}(2004)\citenamefont
  {Sugawara}, \citenamefont {Ebe}, \citenamefont {Hatori}, \citenamefont
  {Ishida}, \citenamefont {Arakawa}, \citenamefont {Akiyama}, \citenamefont
  {Otsubo},\ and\ \citenamefont {Nakata}}]{Sugawara}%
  \BibitemOpen
  \bibfield  {author} {\bibinfo {author} {\bibfnamefont {M.}~\bibnamefont
  {Sugawara}}, \bibinfo {author} {\bibfnamefont {H.}~\bibnamefont {Ebe}},
  \bibinfo {author} {\bibfnamefont {N.}~\bibnamefont {Hatori}}, \bibinfo
  {author} {\bibfnamefont {M.}~\bibnamefont {Ishida}}, \bibinfo {author}
  {\bibfnamefont {Y.}~\bibnamefont {Arakawa}}, \bibinfo {author} {\bibfnamefont
  {T.}~\bibnamefont {Akiyama}}, \bibinfo {author} {\bibfnamefont
  {K.}~\bibnamefont {Otsubo}}, \ and\ \bibinfo {author} {\bibfnamefont
  {Y.}~\bibnamefont {Nakata}},\ }\href {\doibase 10.1103/PhysRevB.69.235332}
  {\bibfield  {journal} {\bibinfo  {journal} {Phys. Rev. B}\ }\textbf {\bibinfo
  {volume} {69}},\ \bibinfo {pages} {235332} (\bibinfo {year}
  {2004})}\BibitemShut {NoStop}%
\bibitem [{\citenamefont {Gurnick}\ and\ \citenamefont
  {DeTemple}(1983)}]{Gurnick}%
  \BibitemOpen
  \bibfield  {author} {\bibinfo {author} {\bibfnamefont {M.}~\bibnamefont
  {Gurnick}}\ and\ \bibinfo {author} {\bibfnamefont {T.}~\bibnamefont
  {DeTemple}},\ }\href@noop {} {\bibfield  {journal} {\bibinfo  {journal} {IEEE
  J. Quantum Elect.}\ }\textbf {\bibinfo {volume} {19}},\ \bibinfo {pages}
  {791} (\bibinfo {year} {1983})}\BibitemShut {NoStop}%
\bibitem [{\citenamefont {Rosencher}\ and\ \citenamefont
  {Bois}(1991)}]{Rosencher}%
  \BibitemOpen
  \bibfield  {author} {\bibinfo {author} {\bibfnamefont {E.}~\bibnamefont
  {Rosencher}}\ and\ \bibinfo {author} {\bibfnamefont {P.}~\bibnamefont
  {Bois}},\ }\href@noop {} {\bibfield  {journal} {\bibinfo  {journal} {Phys.
  Rev. B}\ }\textbf {\bibinfo {volume} {44}},\ \bibinfo {pages} {11315}
  (\bibinfo {year} {1991})}\BibitemShut {NoStop}%
\bibitem [{\citenamefont {Faraon}\ \emph {et~al.}(2008)\citenamefont {Faraon},
  \citenamefont {Fushman}, \citenamefont {Englund}, \citenamefont {Sotltz},
  \citenamefont {Petroff},\ and\ \citenamefont {Vučković}}]{Faraon}%
  \BibitemOpen
  \bibfield  {author} {\bibinfo {author} {\bibfnamefont {A.}~\bibnamefont
  {Faraon}}, \bibinfo {author} {\bibfnamefont {I.}~\bibnamefont {Fushman}},
  \bibinfo {author} {\bibfnamefont {D.}~\bibnamefont {Englund}}, \bibinfo
  {author} {\bibfnamefont {N.}~\bibnamefont {Sotltz}}, \bibinfo {author}
  {\bibfnamefont {P.}~\bibnamefont {Petroff}}, \ and\ \bibinfo {author}
  {\bibfnamefont {J.}~\bibnamefont {Vučković}},\ }\href@noop {} {\bibfield
  {journal} {\bibinfo  {journal} {Nat. Phys.}\ }\textbf {\bibinfo {volume}
  {4}},\ \bibinfo {pages} {859} (\bibinfo {year} {2008})}\BibitemShut {NoStop}%
\bibitem [{\citenamefont {Vahala}(2003)}]{Vahala}%
  \BibitemOpen
  \bibfield  {author} {\bibinfo {author} {\bibfnamefont {K.~J.}\ \bibnamefont
  {Vahala}},\ }\href@noop {} {\bibfield  {journal} {\bibinfo  {journal}
  {Nature}\ }\textbf {\bibinfo {volume} {424}},\ \bibinfo {pages} {1} (\bibinfo
  {year} {2003})}\BibitemShut {NoStop}%
\bibitem [{\citenamefont {Gerry}\ and\ \citenamefont {Knight}(2005)}]{Gerry}%
  \BibitemOpen
  \bibfield  {author} {\bibinfo {author} {\bibfnamefont {C.}~\bibnamefont
  {Gerry}}\ and\ \bibinfo {author} {\bibfnamefont {P.}~\bibnamefont {Knight}},\
  }\href@noop {} {\emph {\bibinfo {title} {Introductory Quantum Optics}}}\
  (\bibinfo  {publisher} {Cambridge University Press},\ \bibinfo {year}
  {2005})\BibitemShut {NoStop}%
\bibitem [{\citenamefont {Lew Yan~Voon}\ and\ \citenamefont
  {Ram-Mohan}(1994)}]{Voon}%
  \BibitemOpen
  \bibfield  {author} {\bibinfo {author} {\bibfnamefont {L.~C.}\ \bibnamefont
  {Lew Yan~Voon}}\ and\ \bibinfo {author} {\bibfnamefont {L.~R.}\ \bibnamefont
  {Ram-Mohan}},\ }\href {\doibase 10.1103/PhysRevB.50.14421} {\bibfield
  {journal} {\bibinfo  {journal} {Phys. Rev. B}\ }\textbf {\bibinfo {volume}
  {50}},\ \bibinfo {pages} {14421} (\bibinfo {year} {1994})}\BibitemShut
  {NoStop}%
\bibitem [{\citenamefont {Andrews}\ \emph {et~al.}(1998)\citenamefont
  {Andrews}, \citenamefont {Pike}, \citenamefont {Sarkar},\ and\ \citenamefont
  {Adlard}}]{Andrews}%
  \BibitemOpen
  \bibfield  {author} {\bibinfo {author} {\bibfnamefont {R.}~\bibnamefont
  {Andrews}}, \bibinfo {author} {\bibfnamefont {E.~R.}\ \bibnamefont {Pike}},
  \bibinfo {author} {\bibfnamefont {S.}~\bibnamefont {Sarkar}}, \ and\ \bibinfo
  {author} {\bibfnamefont {C.}~\bibnamefont {Adlard}},\ }\href@noop {}
  {\bibfield  {journal} {\bibinfo  {journal} {Pure Appl. Opt.}\ }\textbf
  {\bibinfo {volume} {174}},\ \bibinfo {pages} {293} (\bibinfo {year}
  {1998})}\BibitemShut {NoStop}%
\bibitem [{\citenamefont {Drummond}\ and\ \citenamefont
  {Hillery}(2014)}]{Drummond}%
  \BibitemOpen
  \bibfield  {author} {\bibinfo {author} {\bibfnamefont {P.~D.}\ \bibnamefont
  {Drummond}}\ and\ \bibinfo {author} {\bibfnamefont {M.}~\bibnamefont
  {Hillery}},\ }\href@noop {} {\emph {\bibinfo {title} {The Quantum Theory of
  Nonlinear Optics}}}\ (\bibinfo  {publisher} {Cambridge University Press},\
  \bibinfo {year} {2014})\BibitemShut {NoStop}%
\bibitem [{\citenamefont {Kibis}\ \emph {et~al.}(2009)\citenamefont {Kibis},
  \citenamefont {Slepyan}, \citenamefont {Maksimenko},\ and\ \citenamefont
  {Hoffmann}}]{Kibis}%
  \BibitemOpen
  \bibfield  {author} {\bibinfo {author} {\bibfnamefont {O.~V.}\ \bibnamefont
  {Kibis}}, \bibinfo {author} {\bibfnamefont {G.~Y.}\ \bibnamefont {Slepyan}},
  \bibinfo {author} {\bibfnamefont {S.~A.}\ \bibnamefont {Maksimenko}}, \ and\
  \bibinfo {author} {\bibfnamefont {A.}~\bibnamefont {Hoffmann}},\ }\href@noop
  {} {\bibfield  {journal} {\bibinfo  {journal} {Phys. Rev. Lett.}\ }\textbf
  {\bibinfo {volume} {102}},\ \bibinfo {pages} {1} (\bibinfo {year}
  {2009})}\BibitemShut {NoStop}%
\bibitem [{\citenamefont {Jacak}(2000)}]{Jacak}%
  \BibitemOpen
  \bibfield  {author} {\bibinfo {author} {\bibfnamefont {L.}~\bibnamefont
  {Jacak}},\ }\href@noop {} {\bibfield  {journal} {\bibinfo  {journal} {Eur. J.
  Phys.}\ }\textbf {\bibinfo {volume} {21}},\ \bibinfo {pages} {487} (\bibinfo
  {year} {2000})}\BibitemShut {NoStop}%
\bibitem [{\citenamefont {Breuer}\ and\ \citenamefont
  {Petruccione}(2002)}]{Petruccione}%
  \BibitemOpen
  \bibfield  {author} {\bibinfo {author} {\bibfnamefont {H.~P.}\ \bibnamefont
  {Breuer}}\ and\ \bibinfo {author} {\bibfnamefont {F.}~\bibnamefont
  {Petruccione}},\ }\href@noop {} {\emph {\bibinfo {title} {The Theory of Open
  Quantum Systems}}}\ (\bibinfo  {publisher} {Oxford University Press},\
  \bibinfo {year} {2002})\BibitemShut {NoStop}%
\bibitem [{\citenamefont {Bloembergen}\ \emph {et~al.}(1968)\citenamefont
  {Bloembergen}, \citenamefont {Chang}, \citenamefont {Jha},\ and\
  \citenamefont {Lee}}]{Bloembergen}%
  \BibitemOpen
  \bibfield  {author} {\bibinfo {author} {\bibfnamefont {N.}~\bibnamefont
  {Bloembergen}}, \bibinfo {author} {\bibfnamefont {R.~K.}\ \bibnamefont
  {Chang}}, \bibinfo {author} {\bibfnamefont {S.~S.}\ \bibnamefont {Jha}}, \
  and\ \bibinfo {author} {\bibfnamefont {C.~H.}\ \bibnamefont {Lee}},\
  }\href@noop {} {\bibfield  {journal} {\bibinfo  {journal} {P.R.}\ }\textbf
  {\bibinfo {volume} {174}},\ \bibinfo {pages} {813} (\bibinfo {year}
  {1968})}\BibitemShut {NoStop}%
\bibitem [{Note1()}]{Note1}%
  \BibitemOpen
  \bibinfo {note} {See Eq. (3.6.13) in pag. 172 of \cite {Boyd}. A change of
  dummy indexes was made in (\ref {X2class}) to ease comparison.}\BibitemShut
  {Stop}%
\bibitem [{\citenamefont {Savenko}\ \emph {et~al.}(2000)\citenamefont
  {Savenko}, \citenamefont {Kibis},\ and\ \citenamefont {Shelykh}}]{Savenko}%
  \BibitemOpen
  \bibfield  {author} {\bibinfo {author} {\bibfnamefont {I.~G.}\ \bibnamefont
  {Savenko}}, \bibinfo {author} {\bibfnamefont {O.~V.}\ \bibnamefont {Kibis}},
  \ and\ \bibinfo {author} {\bibfnamefont {I.~A.}\ \bibnamefont {Shelykh}},\
  }\href@noop {} {\bibfield  {journal} {\bibinfo  {journal} {P.R.A}\ }\textbf
  {\bibinfo {volume} {85}},\ \bibinfo {pages} {1} (\bibinfo {year}
  {2000})}\BibitemShut {NoStop}%
\bibitem [{\citenamefont {Majumdar}\ \emph {et~al.}(2012)\citenamefont
  {Majumdar}, \citenamefont {Bajcsy},\ and\ \citenamefont
  {Vučković}}]{Majumdar}%
  \BibitemOpen
  \bibfield  {author} {\bibinfo {author} {\bibfnamefont {A.}~\bibnamefont
  {Majumdar}}, \bibinfo {author} {\bibfnamefont {M.}~\bibnamefont {Bajcsy}}, \
  and\ \bibinfo {author} {\bibfnamefont {J.}~\bibnamefont {Vučković}},\
  }\href@noop {} {\bibfield  {journal} {\bibinfo  {journal} {Phys. Rev. A}\
  }\textbf {\bibinfo {volume} {85}},\ \bibinfo {pages} {041801} (\bibinfo
  {year} {2012})}\BibitemShut {NoStop}%
\bibitem [{\citenamefont {Majumdar}(2012)}]{Majumdar_thesis}%
  \BibitemOpen
  \bibfield  {author} {\bibinfo {author} {\bibfnamefont {A.}~\bibnamefont
  {Majumdar}},\ }\emph {\bibinfo {title} {Solid State Cavity Quantum
  Electrodynamics with Quantum Dots Coupled to Photonic Crystal Cavities}},\
  \href@noop {} {Ph.D. thesis},\ \bibinfo  {school} {Stanford University}
  (\bibinfo {year} {2012})\BibitemShut {NoStop}%
\bibitem [{\citenamefont {Garc\'ia}\ \emph {et~al.}(2018)\citenamefont
  {Garc\'ia}, \citenamefont {Rodr\'iguez}, \citenamefont {G\'omez},\ and\
  \citenamefont {Vel\'azquez}}]{SONOP}%
  \BibitemOpen
  \bibfield  {author} {\bibinfo {author} {\bibfnamefont {J.~D.}\ \bibnamefont
  {Garc\'ia}}, \bibinfo {author} {\bibfnamefont {B.~A.}\ \bibnamefont
  {Rodr\'iguez}}, \bibinfo {author} {\bibfnamefont {F.~G.}\ \bibnamefont
  {G\'omez}}, \ and\ \bibinfo {author} {\bibfnamefont {V.~M.}\ \bibnamefont
  {Vel\'azquez}},\ }\href@noop {} {\enquote {\bibinfo {title} {Second order
  nonlinear optical processes in centrosymmetric materials},}\ } (\bibinfo
  {year} {2018}),\ \bibinfo {note} {to be submitted}\BibitemShut {NoStop}%
\end{thebibliography}%
%
%

\end{document}